\documentclass[twoside,11pt]{article}

\usepackage{a4,amsthm,latexsym,amsmath,amssymb,amsfonts,euscript}
\newtheorem{theorem}{{\bf{\small T}{\scriptsize HEOREM}}}[]
\newtheorem{corollary}[]{{\bf{\small C}{\scriptsize OROLLARY}}}
\newtheorem{proposition}[]{{\bf{\small P}{\scriptsize ROPOSITION}}}
\newtheorem{lemma}[]{{\bf{\small L}{\scriptsize EMMA}}}
\newtheorem{remark}[]{{\bf{\small R}{\scriptsize EMARK}}}

\newtheorem{keylemma}[]{{\bf{\small K}{\scriptsize EY-LEMMA}}}

\renewenvironment{proof}[1]
{\noindent{{\bf{\small{ P}{\scriptsize ROOF}}}.}\hspace{0.1cm} #1}
{$\;\qed$\newline}

\def\R{\mathbb R}
\def\N{\mathbb N}
\def\Z{\mathbb Z}
\def\1{{{\mathit 1} \!\!\>\!\! I} }
\def\T{{\mathbf T}}  
\def\L{{\mathbf L}}  
\def\W{{\mathbf W}}

\def\pee{{\mathbb P}}
\def\peer{{\mathbb P}^{{\scriptscriptstyle R}}}
\def\ur{U^{{\scriptscriptstyle R}}}
\def\qee{{\mathbb Q}}
\def\E{\mathbb E}

\def\ent{{\dot{\mathbf S}}}
\def\mep{{\mathbf{MEP}}}

\def\sw{\dot{{\mathcal S}}^{{\scriptscriptstyle W}}_n}
\def\sr{\dot{{\mathcal S}}^{{\scriptscriptstyle H}}_n}

\let\phi=\varphi

\def\limn{\lim_{n\rightarrow\infty}}

\date{}

\begin{document}

\begin{center}
{\sc{\large Testing the irreversibility of a Gibbsian process}}\\
{\sc{\large via hitting and return times}}
\end{center}

\vskip .5truecm

\centerline{(Running title: Entropy production via hitting and return
times)}

\vskip 1.5truecm

\centerline{{\bf J.-R. Chazottes}}
\centerline{Centre de Physique Th{\'e}orique, CNRS UMR 7644}
\centerline{F-91128 Palaiseau Cedex, France}
\centerline{{\tt email~: jeanrene@cpht.polytechnique.fr}}
\centerline{{\bf F. Redig}}
\centerline{Mathematisch Instituut Universiteit Leiden}
\centerline{Niels Bohrweg 1, 2333 CA Leiden, The Netherlands}
\centerline{{\tt email~: redig@math.leidenuniv.nl}}

\vskip 2truecm

\begin{abstract}
We introduce estimators for the entropy production of a Gibbsian
process based on the observation of a single or two typical
trajectories. These estimators are built with adequate hitting and return
times. We then study their convergence and fluctuation properties.
This provides statistical tests for the irreversibility of Gibbsian
processes.

\bigskip


\end{abstract}


\newpage


\section{Introduction}\label{I}
In the theory of non-equilibrium statistical mechanics, the entropy production is a crucial
quantity. Typical for non-equilibrium steady states is the (strict) positivity of the entropy production
which is accompanied by presence of currents and hence breakage of time-reversal symmetry.
In \cite{maes} the entropy production was introduced at the level of trajectories. The idea
is that even for a non-equilibrium system in the steady state, the space-time measure is
still a Gibbs measure and the asymmetric part under time reversal of the Hamiltonian of the space-time Gibbs
measure is the entropy production. Hence, in this formalism, the entropy production is
a trajectory-valued function which measures the degree of irreversibility.
The relative entropy density between the forward and the backward
process is then the {\em mean entropy production}, which is strictly positive if and only if
the process is reversible (i.e., in ``detailed balance'', or
``equilibrium'').
See also \cite{mr} for the relation between strictly
positive mean entropy production and reversibility, and \cite{JQQ} for
a recent account on entropy production in a broader context.

In this point of view, in order to {\em estimate} the entropy production, e.g., in order to
test the reversibility of the process, one needs a way to compute it from trajectories. This is quite similar
to the problem of estimating the entropy of a process. A basic approach
consists in approximating the measure by its empirical version \cite{shields}.
Another particularly useful and simple way of estimating entropy is
via the Ornstein-Weiss theorem \cite{shields,weiss}. The entropy is approximated by the
logarithm of the return time of the first $n$ symbols, divided by $n$. Similarly, relative entropy
density can be estimated using waiting times, see e.g. \cite{konto}.
In this paper we consider Gibbsian processes with values in a finite alphabet, and with
summable modulus of continuity.
We introduce an estimator of the entropy production based on a single trajectory
(we call it the hitting-time estimator) and an estimator based on two independent trajectories
(which we call the waiting-time estimator). For both estimators we obtain consistency and asymptotic
normality, with an asymptotic variance coinciding with that of the entropy production.
Moreover, for the waiting-time estimator we obtain a large deviation principle.
It turns out that its large deviation function has the same symmetry
as in the so-called fluctuation theorem \cite{galco,lebspo,maes}, and in fact coincides
with the large deviation function of the entropy production 
itself in the region where it is finite. This shows that the estimator has also nice properties from
the physical point of view.
The basic technique we use is the exponential law with good control of the error for hitting and waiting
times \cite{miguel,miguelnew}.
This provides us with a precise control of the difference between the estimators and the entropy production.

The rest of the paper is organized as follows. In section 2 we introduce the entropy production in the spirit
of \cite{maes}, see also \cite{JQQ}. In section 3 we introduce the estimators, in section 4 we state their
fluctuation properties and section 5 is devoted to proofs.

\section{Context}\label{C}

We will consider a stationary process $\{X_n:n\in\Z\}$ taking values in a
finite set $A$.
A trajectory of this process, i.e., an element of $A^{\Z}$ will be denoted
by $\omega$.
The space of all trajectories is denoted by $\Omega=A^{\Z}$. For
$\omega\in\Omega$, and
$n\in\Z$, $\theta_n \omega$ is the trajectory defined by $(\theta_n
\omega)_k := \omega_{k+n}$.
A function $f:\Omega\to\R$ is called local if it depends only on finitely
many coordinates
of the trajectory.
A block of length $n$ is a sequence $x_1^n:=x_1\cdots x_n$ of elements of
$A$. The cylinder
$[x_1^n]$ based on $x_1^n$ is the set of $\omega\in\Omega$ such that
$\omega_j=x_j$ for
$j=1,\ldots,n$.

The distribution $\pee$ of the process $\{X_n:n\in\Z\}$ is supposed to be
a translation invariant Gibbs measure with
translation invariant potential $U$. The associated ``energy per site''
$f_U$ is defined as usual:
$$
f_U(\omega):= \sum_{\Lambda\ni 0} \frac{U(\Lambda,\omega)}{|\Lambda|}
$$
where the sum runs over all finite subsets of $\Z$ (containing the
origin).

It is well-known that under mild assumptions \cite{Geo}
there exists a constant $K>0$ such that for all $x_1^n$, all
$\omega\in[x_1^n]$, we have the uniform estimate
\begin{equation}\label{gibbs}
K^{-1} \leq
\frac{\pee([x_1^n])}{\exp(n P(f_U) + \sum_{j=0}^{n-1} f_U(\theta_j
\omega))}
\leq K
\end{equation}
where $P(f_U)$ is the ``pressure'' associated to $U$.

For a block $x_1^n$, its time reverse is denoted by $x_n^1=x_n
x_{n-1}\cdots x_1$.
Similarly, $X_1^n$ denotes the random block $X_1\cdots X_n$ whereas
$X_n^1$ denotes the random block $X_n\cdots X_1$.

For the definition of the entropy production of the process
$\{X_n:n\in\Z\}$, we follow
\cite{maes,mrv}.
We denote by $\peer$ the distribution of the time-reversed process, i.e.,
the distribution
of $\{X_{-n}:n\in\Z\}$.
The entropy production of the process up to time $n$ is defined as
\begin{equation}\label{dracula}
\ent_n(X_1,\ldots,X_n):=\log\frac{\pee([X_1^n])}{\pee([X_n^1])}=
\log\frac{\pee([X_1^n])}{\peer([X_1^n])}\,\cdot
\end{equation}
This random variable is a measure of the irreversibility of the process up
to time $n$.

We recall that the relative entropy density $h(\qee|\pee)$ between a
translation invariant probability
measure $\qee$ on $\Omega$ and $\pee$ is the limit
$$
h(\qee|\pee)=\limn\frac{H_n(\qee|\pee)}{n}
$$
where
$$
H_n(\qee|\pee):=\sum_{x_1^n\in A^n} \qee([x_1^n])
\log\frac{\qee([x_1^n])}{\pee([x_1^n])}\,\cdot
$$
We have the following well-known properties \cite{Geo}:
$$
h(\qee|\pee)=P(f_U)-\int f_U\ d\qee + s(\qee)
$$
where $s(\qee)$ is the entropy density of $\qee$. Moreover,
$h(\qee|\pee)\geq 0$, with equality if and only if $\qee$ is an equilibrium
state for $U$ (variational principle).

Using \eqref{gibbs} and the Ergodic Theorem, it follows immediately that
\begin{equation}\label{chou}
\limn \frac{\ent_n(X_1,\ldots, X_n)}{n}= h(\pee|\peer):=\mep\quad
\pee-\textup{almost surely}\,.
\end{equation}
This quantity is called the {\em mean entropy production}. It is equal to
$0$ if and only
if the process is reversible, i.e., the potential $\ur$ associated to
$\peer$ is physically
equivalent to the potential $U$.

We now precise the classes of potentials for which our results hold.

A first restriction is to assume that $f_U$ has a summable modulus of
continuity, i.e.,
\begin{equation}\label{avion}
\sum_{n\geq 1} \textup{var}_n f_U <\infty
\end{equation}
where
$$
\textup{var}_n f_U := \sup\{ |f_U(\omega)-f_U(\omega')|:
\omega_i=\omega'_i,  \forall |i|\leq n\}\,.
$$
In particular this implies that $\pee$ is the unique Gibbs measure (equilibrium state)
with potential $U$.
It is convenient to work with an $f_U$ which depends only on
``future'' coordinates, that is, only on $\omega_1,\omega_2,\ldots$.
It is indeed proved in \cite{CQ} that if $f_U$ satisfies
\eqref{avion}, then there exists a function
$f_U^+(\omega):=f_U^+(\omega_1,\omega_2,\ldots)$ which is physically
equivalent to $f_U$, i.e., which gives the same Gibbs measure as
$f_U$, and which has also summable variations. ``Physically
equivalent'' means there exists
a measurable function $\kappa=\kappa_U$ and a real constant $C=C_U$
such that $f_U^+ = f_U + \kappa - \kappa\circ \theta + C$. It is
easy to check that \eqref{gibbs} holds with $f_U^+$ in place of
$f_U$ by suitably modifying the constant $K$. Moreover, we can
simplify the notations by assuming that $P(f_U^+)=0$. If it is
not the case, replace $f_U^+$ by the physically equivalent potential
$f_U^+ - P(f_U^+)$. Recapitulating, we obtain that
there exists a constant $K'>0$ such that for all $x_1^n$, all
$\omega\in[x_1^n]$, we have the uniform estimate
\begin{equation}\label{gibbsbis}
K'^{-1} \leq
\frac{\pee([x_1^n])}{\exp(\sum_{j=0}^{n-1} f_U^+(\theta_j \omega))}
\leq K'\,.
\end{equation}
Of course, the same estimate holds for $\peer$ with the obvious
modifications. This immediately gives that there exists some constant
$\tilde{K}>0$ such that
\begin{equation}\label{porc}
-\tilde{K} \leq \ent_n - \sum_{j=0}^{n-1} [(f_U^+ -
  f_{\ur}^+)\circ\theta_j] \leq \tilde{K}
\end{equation}
for all $n\geq 1$.
Using \eqref{chou} and the Ergodic Theorem, we deduce immediately that
$$
\mep = \int (f_U^+ - f_{\ur}^+)\ d\pee\,.
$$

The possibility of working with a ``one-sided'' potential physically
equivalent to the
``two-sided'' one is very important because it will allow us to apply
known results obtained by transfer-operator techniques.

The assumption \eqref{avion} also implies a ``strong mixing'' property
which is needed to
prove our results.
When dealing with central limit asymptotics, we will restrict
ourselves to potentials having exponentially decreasing modulus of
continuity, i.e.,
\begin{equation}\label{bateau}
\exists C>0, 0\leq \eta <1\quad\textup{such that}\quad
\textup{var}_n f_U \leq C \eta^n\quad\forall n\geq 1\,.
\end{equation}
This will allow us to use a result proved in \cite{PP}.
We will precise further these points at the appropriate places.

\begin{remark}
If we assume that
$$
\sum_{\Lambda: \min \Lambda =0} \textup{diam}(\Lambda)
\textup{var}(U(\Lambda,\cdot)) <\infty
$$
where
$\textup{var}(U(\Lambda,\cdot)):=\max(U(\Lambda,\cdot))-\min(U(\Lambda,\cdot))$
this implies \eqref{avion}, see \cite{CQ}.
\end{remark}

\section{Estimators of entropy production based on hitting and return times}\label{wr}

In this section we introduce two estimators based on a single trajectory
or
on two independent trajectories. To define them we have to introduce
hitting times.

The hitting time of a cylinder $[x_1^n]$ is defined as
$$
\T_{x_1^n}(\omega)
:=\inf\{k\geq 1: \theta_k \omega\in [x_1^n] \}\,.
$$
For the sake of convenience, we introduce the notations
$$
\T^+_n(\omega):=\T_{\omega_1^n}(\omega)
\quad\textup{and}\quad
\T^-_n(\omega):=\T_{\omega_n^1}(\omega)\,.
$$

The hitting-time estimator $\sr(\omega)$ of the entropy production is
defined as
$$
\sr(\omega):= \log\frac{\T^-_n(\omega)}{\T^+_n(\omega)}\,\cdot
$$
In words, this is the difference of the logarithms of the first time at which we observe
the first $n$ symbols in reversed order in the trajectory and the first return time of the
first $n$ symbols.
It will follow from our analysis that {\em typically},
$\T^-_n\gg \T^+_n$ if the process is not reversible.
Hence our hitting-time estimator of the entropy production will be
typically positive.

The waiting-time estimator $\sw(\omega,\omega')$ of the entropy production
is based on
two trajectories $\omega, \omega'$ chosen {\em independently} of one
another according to
$\pee$. We introduce the following convenient notations:
$$
\W^+_n(\omega,\omega'):=\T_{\omega_1^n}(\omega')\quad\textup{and}\quad
\W^-_n(\omega,\omega'):=\T_{\omega_n^1}(\omega')\,.
$$
The waiting-time estimator is then defined as
$$
\sw(\omega,\omega'):=
\log\frac{\W^-_n(\omega,\omega')}{\W^+_n(\omega,\omega')}\,\cdot
$$
The main motivation to introduce this alternative estimator is that we
will obtain
a better control of its large deviation properties.

\begin{remark}
We can define two other estimators based on the so-called matching
times \cite{konto}. They are in some sense the ``duals'' of the above
estimators.
To introduce the ``dual" of the hitting-time estimator, consider the first
$n$ symbols $x_1,\ldots x_n$ of the process and define
\[
\L^+_n = \min \{ k\leq n: \mbox{the word}\ x_1^k\ \mbox{does not reappear in} \ x_1^n\}
\]
and
\[
\L^-_n = \min \{ k\leq n: \mbox{the reversed word}\ x_k^1\ \mbox{does not reappear in} \ x_1^n\}
\]
Then the estimator of the entropy production dual to the hitting-time estimator is given by
$\log(\L^+_n/\L^+_-)$.

The advantage of these estimators is that they are
based on a trajectory of finite length $n$. However, all the asymptotic fluctuation properties
of these estimators can be derived from the ones of the present paper by the duality relations.
So we do not study them in detail in this paper.
\end{remark}

\section{Convergence and fluctuations of the estimators}\label{MR}

We now state our results on consistency and asymptotic normality for the
estimators we just introduced, as well as large deviation properties for estimators based on
two independent trajectories. Recall that $\mep$ is the mean entropy production, see
\eqref{chou}.

\subsection{Almost-sure approximation and consistency}

The following theorem provides an almost-sure approximation of $\ent_n$, the
entropy production up to time $n$  (see \eqref{dracula}),
by both the return-time and the waiting-time estimators.

\begin{theorem}\label{thm1}
Assume that \eqref{avion} holds. Then there exists a constant $C=C(\pee)>0$ such that
\begin{enumerate}
\item Eventually $\pee$-almost surely
$$
-C\log n\leq
\sr - \ent_n
\leq C\log n\,;
$$

\item Eventually $\pee\!\times\!\pee$-almost surely
$$
-C\log n\leq
\sw - \ent_n
\leq C\log n\,.
$$
\end{enumerate}
\end{theorem}

Using the previous theorem and \eqref{chou}, we immediately obtain the
following corollary
establishing the consistency of our entropy production estimators.

\begin{corollary}
We have the following almost-sure convergences:
\begin{enumerate}
\item $\pee$-almost surely
$$
\limn\frac{\sr}{n}= \mep\,;
$$
\item $\pee\!\times\!\pee$-almost surely
$$
\limn\frac{\sw}{n}= \mep\, .
$$
\end{enumerate}
\end{corollary}

\subsection{Asymptotic normality}

The expectation with respect to $\pee$ is denoted by $\E$.
Let
\begin{equation}\label{variance}
\sigma^2:=\sum_{\ell\geq 1} \left[
\E((f_U^+ -f_{\ur}^+)\cdot (f_U^+ -f_{\ur}^+)\circ
\theta_\ell)-(\E(f_U^+ - f_{\ur}^+))^2\right]\,.
\end{equation}
It can be showed that $\sigma^2<\infty$ if \eqref{bateau} holds.
It is well-known that $\sigma^2>0$ unless $U$ is physically equivalent to
$\ur$,
i.e., $f_U^+ - f_{\ur}^+ $ is a co-boundary, which in turn is equivalent
with $\pee=\peer$,
i.e., the process is reversible.
For more details on this, we refer to \cite{PP}.

\begin{theorem}\label{pouac}
Assume that \eqref{bateau} holds. Then
we have the following central limit asymptotics:
\begin{enumerate}
\item For the hitting-time estimator
$$
\frac{\sr - n \mep}{\sqrt{n}}\to
\mathcal{N}(0,\sigma^2)\,,\textup{as}\;n\to\infty
$$
in $\pee$-distribution.
\item For the waiting-time estimator
$$
\frac{\sw - n \mep}{\sqrt{n}}\to
\mathcal{N}(0,\sigma^2)\,,\textup{as}\;n\to\infty
$$
in $\pee\!\times\!\pee$-distribution.
\end{enumerate}
Moreover,
\begin{equation}\label{varvar}
\limn\frac{\textup{Var}(\sr)}{n}=\limn\frac{\textup{Var}(\sw)}{n}=
\sigma^2
\end{equation}
where $\textup{Var}$ denotes the variance.
\end{theorem}

\begin{remark}
Using the results of \cite{KMS}, we could extend the previous theorem
to potentials with a modulus of continuity decreasing polynomially,
i.e., like $1/n^{\alpha}$ for $\alpha>0$ large enough.
\end{remark}

\subsection{Large deviations}

Our goal is to analyze the deviations of order one of $\sw/n$ around the mean
entropy production $\mep$. To this end, we introduce the following ``free-energy-like'' function,
which is nothing but the scaled-cumulant generating function for the process $(\sw)$:
$$
\mathcal{W}_U(p):=\limn\frac{1}{n}\log
\E_{\pee\!\times\!\pee}\left(e^{p\sw}\right)\, , \; p\in\R
$$
provided the limit exists.
On another hand, define the scaled cumulant generating function for the
process $(\ent_n)$ as:
$$
\mathcal{E}_U(p):=\limn\frac{1}{n}\log \E_{\pee}\left(e^{p\ent_n}\right)\,
, p\in\R\,.
$$
It is easy to deduce from \eqref{gibbsbis} that
$$
\mathcal{E}_U(p)=P(-p f_{\ur}^+ + (1+p)f_U^+)\ ,\;\forall p\in\R\,.
$$
From this formula one immediately sees that
$$
\mathcal{E}_U(-1-p)
=
\mathcal{E}_{\ur}(p)\,.
$$
On another hand, it is obvious from the definition of $\ent_n$ that
$$
\mathcal{E}_U(p)
=
\mathcal{E}_{\ur}(p)\,.
$$
Hence
$$
\mathcal{E}_U(-1-p)
=
\mathcal{E}_{U}(p)
$$
which is a version of the Gallavotti-Cohen fluctuation theorem, see
\cite{galco}, \cite{lebspo}, \cite{maes}.

Notice that $\mathcal{E}_U\equiv 0$ if $U$ is physically equivalent
to $\ur$.

We now state a large deviation result for $\ent_n$. 
Let ${\mathcal I}_U$ be the Legendre transform of ${\mathcal E}_U$, i.e.,
$$
{\mathcal I}_U(q)= \sup_{p\in\R} \left(pq - {\mathcal E}_U(p)\right) \,.
$$
Then we have

\begin{proposition}\label{macbeth}
Assume that \eqref{avion} holds and that the process $(X_n)$ is not reversible (i.e., that
$U$ is not physically equivalent to $\ur$). Then the function $p\mapsto {\mathcal E}_U(p)$
is continuously differentiable and strictly convex. Moreover,
there exists an open interval $(\underline{q}, \overline{q})$ such that, for every interval $J$
with $J\cap (\underline{q}, \overline{q})\neq \emptyset$
$$
\limn \frac{1}{n}\log \pee\left\{\frac{\ent_n(X_1,\ldots,X_n)}{n} \in J \right\}=
-\inf_{q\in J\cap (\underline{q}, \overline{q})} {\mathcal I}_U(q)\,.
$$
\end{proposition}

The interest of this result lies in its formulation adapted to our
context and convenient to state the next result, the main one of this
section. In essence such kind of result appears, e.g., in \cite{MV}.

\begin{theorem}\label{pouic}
If assumption \eqref{avion} holds 
then we have
\begin{equation}
\mathcal{W}_U(p)=
\left\{
\begin{array}{l}
\mathcal{E}_U(p)\quad\textup{if}\;-1<p<1\\
+\infty \quad\textup{otherwise}\,.
\end{array}\label{boulgakov}
\right.
\end{equation}
In particular, if the process $(X_n)$ is not reversible (i.e., $U$ is not physically equivalent to $\ur$) then 
$\sw$ and $\ent_n$ have the same large deviations in the
open interval $(c_-,c_+)$, with $c_-:=\lim_{p\to -1}\mathcal{E}_U'(p)<0$
and $c_+:=\lim_{p\to 1}\mathcal{E}_U'(p)>0$: For every interval $J$
with $J\cap (c_-, c_+)\neq \emptyset$
\begin{equation}\label{pelleas}
\limn \frac{1}{n}\log \pee\left\{\frac{\sw}{n} \in J \right\}=
-\inf_{q\in J\cap (c_-, c_+)} {\mathcal I}_U(q)\,.
\end{equation}
\end{theorem}

\bigskip

It is easy to check that $\mep\in (c_-,c_+)$. Indeed ${\mathcal E}_U'(0)=\mep$ (one
uses differentiability and convexity to prove that).

The next proposition highlights the symmetry properties of $\mathcal{W}$.
We write explicitly the dependence of $\mathcal{W}$ on the potential $U$.

\begin{proposition}\label{symmetry}
Under  assumption \eqref{avion} we have the following identities
\begin{enumerate}
\item For all $-1<p\leq 0$, we have
$$
\mathcal{W}_{U}(-1-p)=
\mathcal{W}_{\ur}(p) =
\mathcal{W}_U(p)=
\mathcal{W}_{\ur}(-1-p)\,.
$$
\item For all $p\in(-1,1)$, we have
$$
\mathcal{W}_{U}(p) =
\mathcal{W}_{\ur}(p)\,.
$$
\end{enumerate}
\end{proposition}

\begin{remark}
One may ask why we did not study the large deviations of $\sr$, the
hitting-time estimator. Indeed, the analysis of the corresponding
scaled cumulant generating function is made more complicated due
to the effect of ``too soon'' recurrent cylinders. We shall not
detail more on this. Following the approach of \cite{CGS}, we can
obtain a partial counterpart of Theorem \ref{pouic} for $\sr$ :
its scaled cumulant generating function coincides with
$\mathcal{E}_U(p)$ but only in an {\em implicit} interval
$[\tilde{c}_-,\tilde{c}_+]$, where $\tilde{c}_-<0$ and
$\tilde{c}_+>0$.
\end{remark}

\section{Proofs}\label{proofs}

\subsection{Key lemmas}

The following results are the main tools to derive our results.

\begin{keylemma}\label{MKL}
Assume that $\pee$ is a translation invariant Gibbs measure such that
\eqref{avion} holds.
Then there exist strictly positive constants $c,C,\rho_1,\rho_2$, with
$\rho_1\leq \rho_2$,
such that for all $n\in\N$, all cylinders $[a_1^n]$ and all $t>0$
there exists $\rho(a_1^n)\in[\rho_1,\rho_2]$ such that
\begin{equation}\label{strong-approximation}
\Big\vert \pee\{\T_{a_1^n}\pee([a_1^n])>t\}- e^{-\rho(a_1^n)t}\Big\vert
\leq C e^{-c n} e^{-\rho(a_1^n)t}\,.
\end{equation}
\end{keylemma}

\begin{proof}
In \cite{miguel}, the author proved this result
under the assumption that the process is $\psi$-mixing.
Besides, it is proved in \cite{walters}  that if $f_U^+$ has summable
variations, then the process $(X_n)$ is $\psi$-mixing. (This can
be read off the proof of Theorem 3.2 in \cite{walters}.)
\end{proof}

The next lemma will be crucial to control certain moments. This is a
rewriting of Lemma 9 in \cite{miguel}.
\begin{lemma}\label{tarte}
For all cylinder $[a_1^n]$, all $t$ such that $t \leq 1/2$, we have
$$
1-e^{-\rho_1 t} \leq \pee\{\T_{a_1^n}\pee([a_1^n])\leq t\}\leq 1-
e^{-\rho_2 t}
$$
where $\rho_1,\rho_2$ are the constants of Key-lemma \ref{MKL}.
\end{lemma}

We now state the analog to Key-lemma \ref{MKL} for return times.
To do so, we need to define the set of $n$-cylinders with ``internal
periodicity'' $k\leq n$:
$$
\mathcal{S}_k(n):=\{[a_1^n]: \min\{j\in\{1,...,n\}: [a_1^n]\cap
\theta_{j}[a_1^n]\neq\emptyset\}=k \}\,.
$$
Notice that the set of $n$-cylinders can be written as $\bigcup_{1\leq
p\leq n} \mathcal{S}_k(n)$.

\begin{keylemma}\label{MKLbis}
Assume that $\pee$ is a translation invariant Gibbs measure such that
\eqref{avion} holds.
Then there exist strictly positive constants $c,c',C$
such that for any $n\in\N$, any $k\in\{1,...,n\}$,\
any cylinder $[a_1^n] \in \mathcal{S}_k(n)$, one has for all $t\geq k$
\begin{equation}
\Big\vert \pee\big\{\omega:\T_{a_1^n}(\omega)\pee([a_1^n]) >t\big\vert\
[a_1^n]\big\}
- \zeta(a_1^n) \exp(-\zeta(a_1^n)t)\Big\vert
\leq
C \ e^{-c n}\ e^{-c' t}
\end{equation}
where $\zeta(a_1^n)$ is such that $|\ \zeta(a_1^n)-\rho(a_1^n)|\leq D e^{-c n}$,
for some $D>0$. The parameter $\rho(a_1^n)$ is defined in Key-lemma \ref{MKL}.
Moreover,
$$
\pee\{\omega:\T_{a_1^n}(\omega)>t\ | \ [a_1^n]\}=1\quad \textup{for
all}\quad t<k\, .
$$
\end{keylemma}

\begin{proof}
This Key-lemma is a rewriting of \cite[Section 6]{miguelnew}. As for the
previous Key-lemma, the assumption is that the process is $\psi$-mixing.
\end{proof}

\subsection{Proof of Theorem \ref{thm1}}

Let us start with the proof of the second statement of the theorem. We
shall prove
that eventually $\pee\!\times\!\pee$-almost surely
\begin{equation}\label{pizza}
-C_1\log n\leq \log(\W_n^+(\omega,\omega') \pee([\omega_1^n])\leq \log C_1 +\log\log n
\end{equation}
for some $C_1>0$.
It will be clear that by the same reasoning we will also have
that eventually $\pee\!\times\!\pee$-almost surely
\begin{equation}
-C_2\log n\leq \log(\W_n^{-}(\omega,\omega') \pee([\omega_n^1])\leq
\log C_2 +\log\log n
\end{equation}
for some $C_2>0$.
Putting together these two results  immediately gives the
statement 2 of the theorem.

We first prove the upper bound in \eqref{pizza}. We want to find a
summable upper-bound to
$$
\pee\!\times\!\pee\{\log(\W_n^+ \pee([x_1^n]))> \log t\}=
$$
\begin{equation}\label{split}
\sum_{x_1^n}\pee([x_1^n])\
\pee\left\{\log(\T_{x_1^n} \pee([x_1^n]))> \log t \right\}
\end{equation}
where $t$ will be a suitable function of $n$.
We apply Key-lemma \ref{MKL} to get for all $t>0$
$$
\pee\!\times\!\pee\{\log(\W_n^+ \pee([x_1^n]))> \log t\}
\leq C e^{-cn}+ e^{-\rho_1 t}\, .
$$
Take $t=t_n= \log n^{\alpha_1}$, $\alpha_1>0$, to get
$$
\pee\!\times\!\pee\{\log(\W_n^+ \pee([x_1^n]))> \log\log n^\alpha\}
\leq  C e^{-cn} + \frac{1}{n^{\rho_1 \alpha_1}}\,.
$$
By the Borel-Cantelli Lemma we get
$$
\log(\W_n^+ \pee([a_1^n]))\leq \log\log n^{\alpha_1}
$$
eventually $\pee\!\times\!\pee$-almost surely provided that
$\alpha_1\rho_1 >1$.

To obtain the lower bound in \eqref{pizza}, we have, by Key-lemma
\ref{MKL}
$$
\pee\!\times\!\pee\{\log(\W^+_n \pee([x_1^n]))\leq \log t\}
\leq C e^{-cn}+ 1-e^{-\rho_2 t}\leq C e^{-cn}+\rho_2 t
$$
for all $t>0$.
Choose $t=t_n=n^{-\alpha_2}$, $\alpha_2>1$ and apply the Borel-Cantelli
Lemma to get
$$
\log(\W_n^+ \pee([x_1^n]))> -\alpha_2\log n
$$
eventually $\pee\!\times\!\pee$-almost surely.

Let us now prove the first statement of the
theorem. The proof is very similar except we have to deal with
``bad'' cylinders and use Key-lemma \ref{MKLbis}. We will only establish
that eventually $\pee$-almost
surely the inequality
\begin{equation}\label{bof}
-C_1\log n\leq \log(\T_n^+(\omega) \pee([\omega_1^n])\leq \log C_1 +\log\log n
\end{equation}
for some $C_1>0$. The analogous inequality for $\T^-(\omega)$ is
obtained as above (i.e., using Key-lemma \ref{MKL}).
We have the decomposition
$$
\pee\{\omega:\log(\T_n^+(\omega) \pee([\omega_1^n]))> \log t\}=
$$
$$
\sum_{k=1}^n \sum_{x_1^n\in \mathcal{S}_k(n)}\pee([x_1^n])\
\pee\left\{\omega:\log(\T_{x_1^n}(\omega) \pee([x_1^n])> \log t \ | \
[x_1^n]\right\}
$$
where $\mathcal{S}_k(n)$ is defined just before we state Key-lemma
\ref{MKLbis}.
For all $t\geq k \pee([x_1^n])$ and $n$ large enough, we get using
Key-lemma \ref{MKLbis}
$$
\pee\left\{\log(\T_{x_1^n} \pee([x_1^n])> \log t \ | \ [x_1^n])\right\}\leq
(\rho_2 + D) e^{-\frac{\rho_1}{2}t} + C e^{-cn}
$$
where we used the fact that if $n$ is large enough,
$\rho_1/2 \leq \rho_1 - D e^{-c n} \leq \zeta(a_1^n) \leq \rho_2 +
D$. We now choose $t=t_n=\log n^{\alpha_1}$, $\alpha_1>0$. If $n$ is
large enough, then $t_n \geq k \pee([a_1^n])$.
This is because we have the uniform estimate $\pee([a_1^n])\leq
e^{-G n}$, for some $G>0$, since $\pee$ is a Gibbs measure.
Hence we obtain
$$
\pee\{\log(\T_n^+ \pee([\omega_1^n]))> \log\log n^{\alpha_1}\} \leq
\frac{\rho_2 + D}{n^{\alpha_1 \rho_1/2}} + C e^{-cn}
$$
which is summable provided that $\alpha_1 \rho_1/2 >1$. The
Borel-Cantelli Lemma then gives the upper-bound in \eqref{bof}.
The lower-bound is obtained as for the waiting-time estimator but
using Key-lemma \ref{MKLbis}. \hfill $\qed$

\subsection{Proof of Theorem \ref{pouac}}

Let us prove the second statement of the theorem and that
$\limn\frac{\textup{Var}(\sw)}{n}= \sigma^2$.
For this it is enough to prove that
\begin{equation}\label{bong}
\limn\frac{1}{n}\int \left(\sw - \ent_n \right)^2 d\pee\!\times\!\pee=0\,.
\end{equation}
Indeed, proving \eqref{bong} implies, on one hand, that
$$
\limn\frac{\textup{Var}(\ent_n)}{n}=\limn\frac{\textup{Var}(\sw)}{n}\,\cdot
$$
On the other hand, it also implies that
$(\sw-n\mep)/\sqrt{n}$ converges in law to the normal
$\mathcal{N}(0,\sigma^2)$
if, and only if, $(\ent_n-n\mep)/\sqrt{n}$ converges in law to the same law.

Now it is obvious from \eqref{porc} that
$$
\frac{\ent_n - \sum_{j=0}^{n-1} [(f_U^+ -
  f_{\ur}^+)\circ\theta_j]}{\sqrt{n}}\to 0\quad
\pee-\textup{almost-surely}\,.
$$
By applying a result of \cite{PP}, we obtain that
$$
\frac{\sum_{j=0}^{n-1} [(f_U^+ -
  f_{\ur}^+)\circ\theta_j] - n\mep }{\sqrt{n}}\stackrel{\textup{in
law}}{\longrightarrow}
  \mathcal{N}(0,\sigma^2)\,.
$$
Since we have the formula (see \cite{PP})
$$
\sigma^2=\limn \frac{1}{n}\int \big(
\sum_{j=0}^{n-1} [(f_U^+ - f_{\ur}^+)\circ\theta_j] - n\mep\big)^2\ d\pee
$$
it is obvious by \eqref{porc} that
$$
\limn\frac{\textup{Var}(\ent_n)}{n}=\sigma^2\,.
$$

Therefore we have reduced the statements of the theorem about $\sw$ to
proving \eqref{bong}.
By definition we have
$$
\int \left(\sw - \ent_n \right)^2 d\pee\!\times\!\pee=
\sum_{x_{1}^{n}} \pee([x_1^n]) \int \left[
\log(\T_{n}^{-}\pee([x_n^1]))-\log(\T_{n}^{+}\pee([x_1^n]))
\right]^2 d\pee\ .
$$
Let us now prove that the integral in the rhs is bounded above by a positive
number
independent of $n$, implying immediately \eqref{bong}. To prove this
assertion,
it is sufficient to prove that
\begin{equation}\label{bing}
\int \left[\log(\T_{n}^{+}\pee([x_1^n]))\right]^2 d\pee \leq D_1,\quad
\int \left[\log(\T_{n}^{-}\pee([x_n^1]))\right]^2 d\pee \leq D_2\quad
\end{equation}
where $D_1, D_2>0$ are independent of $n$.
We only prove the first inequality since the other one is proved in
exactly the
same way.

We have the following identities:
$$
\int \left[\log(\T_{n}^{+}\pee([x_1^n]))\right]^2 d\pee = \int_0^\infty
\pee\left([\log(\T_{n}^{+}\pee([x_1^n]))]^2 >t \right) dt =
$$
$$
2 \int_1^\infty \pee\left( \T_{n}^{+}\pee([x_1^n]) >t \right)
\frac{\log t}{t}\ dt
+
2 \int_0^1 \pee\left( \T_{n}^{+}\pee([x_1^n]) <t \right)
\frac{-\log t}{t}\ dt=
$$
$$
2 \int_1^\infty \pee\left( \T_{x_1^n}\pee([x_1^n]) >t \right)
\frac{\log t}{t}\ dt
+
2 \int_0^1 \pee\left( \T_{x_1^n}\pee([x_1^n]) <t \right)
\frac{-\log t}{t}\ dt =: \textup{I}\, + \, \textup{II}\,.
$$
Now we use Key-lemma \ref{MKL} and get
$$
\pee\left( \T_{x_1^n}\pee([x_1^n]) >t \right) \leq (1+C) e^{-\rho_1 t},\;\forall n\geq 1\,.
$$
Therefore
$$
\textup{I}\leq 2(1+C) \int_1^\infty \frac{\log t}{t} \ e^{-\rho_1 t} \ dt=:D_1'<\infty\,.
$$
For the integral II, we have the following estimates
$$
\textup{II} = 2 \left(\int_0^{\frac{1}{2}} +\int_{\frac{1}{2}}^1
\right)
\pee\left( \T_{x_1^n}\pee([x_1^n]) <t \right)
\frac{-\log t}{t}\ dt \leq
$$
$$
\int_0^{\frac{1}{2}}\frac{-\log t}{t}\ (1-e^{-\rho_2 t})\ dt 
+
\int_{\frac{1}{2}}^1 \frac{-\log t}{t}\ dt \leq
$$
$$
-\rho_2 \int_0^{\frac{1}{2}} \log t\ dt - \int_{\frac{1}{2}}^1 \frac{\log
  t}{t}\ dt := D_1''<\infty
$$
where we used Lemma \ref{tarte} to bound the first integral.

This finishes the proof for the waiting-time estimator. Concerning the
hitting-time estimator, we leave the proof to the reader. It is very
similar to the previous one except that one has to use Key-lemma
\ref{MKLbis}. \hfill $\qed$

\subsection{Proof of Proposition \ref{macbeth}}

The proof is an application of G\"artner-Ellis theorem \cite{ellis}. In particular we have
to check that the function $p\mapsto \mathcal{E}_U(p)$ is continuously differentiable and strictly convex
under assumption \eqref{avion}. 
The strict convexity follows from the assumption that 
the process is not reversible. As already mentioned above, this amounts to
requiring that $U$ is not physically equivalent to $\ur$, i.e., that $f_U^+ - f_{\ur}^+$ is not a
co-boundary. The open interval $(\underline{q},\overline{q})$ is defined by
$\underline{q}=\inf_{q\in\R}=\lim_{p\to-\infty} {\mathcal E}_U'(p)$ and
$\overline{q}=\sup_{q\in\R}=\lim_{p\to+\infty} {\mathcal E}_U'(p)$. These limits exist by convexity
arguments.
We refer to \cite{israel} from which one can deduce these classical facts on differentiability and
convexity of the pressure function. \hfill $\qed$

\subsection{Proof of Theorem \ref{pouic}}

We prove formula \eqref{boulgakov}.
We first deal with $0<p<1$. The case $-1<p<0$ is obtained by a similar
reasoning, so we omit the proof. The case $p=0$ is trivial.

We observe that
$$
\E_{\pee\!\times\!\pee}\left(e^{p\sw}\right)=
\sum_{x_1^n} \pee([x_1^n])^{p+1} \pee([x_n^1])^{-p}\
\E_\pee\left[\left(\frac{Y_n}{Z_n}\right)^p\right]
$$
where $Y_n:=\T_{x_n^1} \pee([x_n^1])$,
$Z_n:=\T_{x_1^n} \pee([x_1^n])$.
We then have
$$
\E_\pee\left[\left(\frac{Y_n}{Z_n}\right)^p\right] =
\int_0^\infty dy \int_0^\infty dz\
\left(\frac{y}{z}\right)^p
\ \pee\{Y_n\in dy,Z_n\in dz\}
$$
\begin{equation}
=  \int_0^1 dy\int_0^1 dz\
\left(\frac{y}{z}\right)^p
\ \pee\{Y_n\in dy,Z_n\in dz\}
+
\int_1^\infty dy \int_1^\infty dz\
\left(\frac{y}{z}\right)^p \ \pee\{Y_n\in dy,Z_n\in dz\}
\label{cigare}
\end{equation}

We obtain the obvious upper bound
\begin{eqnarray}
\nonumber
\eqref{cigare} & \leq &
\int_0^1 dy\int_0^1 dz\
\frac{1}{z^p}
\ \pee\{Y_n\in dy,Z_n\in dz\}
+
\int_1^\infty dy \int_1^\infty dz\
y^p \ \pee\{Y_n\in dy,Z_n\in dz\}
\\
& \leq & \E_{\pee}\left(\frac{1}{Z_n^p}\right) +\E_{\pee}(Y_n^p)\,.
\label{briquet}
\end{eqnarray}

We get easily the lower bound
\begin{eqnarray}
\nonumber
\eqref{cigare} & \geq &
\int_1^\infty dy \int_1^\infty dz\
\frac{1}{z^p} \ \pee\{Y_n\in dy,Z_n\in dz\}
\\
& \geq & \E_{\pee}\left(\frac{1}{Z_n^p} \1\{Z_n\geq 1\} \right)
\label{allu}
\end{eqnarray}
where $\1\{\cdot\}$ denotes the indicator function.

Proving Theorem \ref{pouic} for $0<p<1$ is thus reduced to proving that
the rhs in \eqref{briquet}
is bounded above by a positive number independent of $n$, and that the rhs
in \eqref{allu} is bounded below
by a positive number independent of $n$.

Let us start with an upper bound for $\E_{\pee}(Y_n^p)$. We have
$$
\E_{\pee}(Y_n^p)= p \int_0^\infty y^{p-1} \pee\{\T_{x_n^1}
\pee([x_n^1])>y\}\ dy\,.
$$
By using Key-lemma \ref{MKL} with $a_1^n=x_n^1$, we obviously have
$\pee\{\T_{x_n^1}
\pee([x_n^1])>y\}<A e^{-By}$ for some $A,B>0$.

Let us now upper-bound $\E_{\pee}\left(\frac{1}{Z_n^p}\right)$.
We have
$$
\E_{\pee}\left(\frac{1}{Z_n^p}\right)=
|p| \left(\int_0^{\frac{1}{2}}+\int_{\frac{1}{2}}^\infty\right)
 z^{-|p|-1} \pee\{\T_{x_1^n} \pee([x_1^n])\leq z\}\ dz\,.
$$
The integral from $\frac{1}{2}$ to $\infty$ is bounded above by
$\int_{\frac{1}{2}}^\infty z^{-|p|-1} dz<\infty$.
To bound the other integral we use Lemma \ref{tarte}:
$$
\int_0^{\frac{1}{2}}z^{-|p|-1} \pee\{\T_{x_1^n} \pee([x_1^n])\leq z\}\
dz\leq \int_0^{\frac{1}{2}} \frac{1-e^{-\rho_2 z}}{z^{|p|+1}} dz
<\infty\,.
$$

We now estimate from below $\E_{\pee}(Y_n^p \1\{Y_n\geq 1\})$.
We have
$$
\E_{\pee}(Y_n^p \1\{Y_n\geq 1\})=
|p| \int_1^\infty  y^{-|p|-1} \pee\{\T_{x_n^1} \pee([x_n^1])\leq y\}\
dy\,.
$$
By Key-lemma \ref{MKL} with $a_1^n=x_1^n$ we have
$$
\pee\{\T_{x_n^1} \pee([x_n^1])\leq y\} \geq 1-(1+C e^{-cn}) e^{-\rho_1
y}\,.
$$
Observe that $1-(1+C e^{-cn}) e^{-\rho_1 y} \geq 1-(1+C e^{-cn})
e^{-\rho_1}$
for all $y\geq 1$ and for all $n\geq 1$.
Therefore
$$
\E_{\pee}(Y_n^p \1\{Y_n\geq 1\}) \geq 1-(1+C e^{-cn}) e^{-\rho_1}>0
$$
provided that $n$ is large enough.

Recapitulating, we proved that for all $0<p<1$ and all $n$ large enough
$$
E^{-1}\leq \E_\pee\left[\left(\frac{Y_n}{Z_n}\right)^p\right] \leq E
$$
for some $E>0$ independent of $n$ and $x_1^n$. Hence, for all $0<p<1$,
we get
$$
\limn\frac{1}{n}\log \E_{\pee\!\times\!\pee}\left(e^{p\sw}\right)=
\limn\frac{1}{n}\log \sum_{x_1^n} \pee([x_1^n])^{p+1}
\pee([x_n^1])^{-p}=
\mathcal{E}_U(p)\,.
$$
The last equality follows obviously from \eqref{gibbsbis} and
\eqref{porc}.

\bigskip

We now turn to the case $|p|\geq 1$. We only deal with the case
$p\geq 1$ since the case $p\leq -1$ is obtained by the same reasoning.
We have
\begin{eqnarray}
\E_\pee\left[\left(\frac{Y_n}{Z_n}\right)^p\right] & \geq &
p \int_0^1 \frac{1}{y^{p+1}}\ \pee\{\T_{x_1^n} \pee([x_1^n])\leq y\}\ dy\\
& \geq &
p \int_0^1 \frac{1}{y^{p+1}}\ (1-(1+C e^{-cn}) e^{-\rho_1 y})\ dy\\
& = &+ \infty
\end{eqnarray}
for $n$ large enough and where we used Key-lemma \ref{MKL} to get the
second inequality.

To prove \eqref{pelleas}, we apply a variant of G\"artner-Ellis theorem found in
\cite{PS}. To this end, we use formula \eqref{boulgakov} and the differentiability/convexity
properties of the function $p\mapsto {\mathcal E}_U(p)$. We have to restrict to
the interval $(c_-,c_+)$ where ${\mathcal W}_U$ is finite and coincides with ${\mathcal E}_U$.
\hfill $\qed$


\end{document}